\documentclass[%
 reprint,
 longbibliography,
superscriptaddress,
 amsmath,amssymb,
 aps,
 10pt,
prl
]{revtex4-2}

\usepackage{graphicx}
\usepackage{dcolumn}
\usepackage{bm}
\usepackage{amssymb}
\usepackage{amsmath}
\usepackage{comment}
\usepackage{lineno}

\usepackage{hyperref}
\hypersetup{colorlinks=true, linkcolor=blue, citecolor=blue, urlcolor=blue,}

\usepackage{orcidlink}

\begin{document}

\title{X-ray-induced quenching of the $^{229}$Th clock isomer in CaF$_2$}


\author{Ming Guan\orcidlink{0009-0000-3563-2516}}
\thanks{guanming-s@s.okayama-u.ac.jp}
\affiliation{Research Institute for Interdisciplinary Science, Okayama University, Okayama 700-8530, Japan}
\author{Michael Bartokos\orcidlink{0009-0004-7610-9122}}
\affiliation{Institute for Atomic and Subatomic Physics, Atominstitut, TU Wien, Vienna 1020, Austria}
\author{Kjeld Beeks\orcidlink{0000-0002-8707-6723}}
\affiliation{Institute for Atomic and Subatomic Physics, Atominstitut, TU Wien, Vienna 1020, Austria}
\author{Hiroyuki Fujimoto\orcidlink{https://orcid.org/0000-0001-6656-7147}}
\affiliation{National Institute of Advanced Industrial Science and Technology (AIST),   Ibaraki 305-8563, Japan}
\author{Yuta Fukunaga}
\affiliation{Research Institute for Interdisciplinary Science, Okayama University, Okayama 700-8530, Japan}
\author{Hiromitsu Haba\orcidlink{https://orcid.org/0000-0002-0170-8305}}
\affiliation{RIKEN, 2-1 Hirosawa, Wako, Saitama 351-0198, Japan}
\author{Takahiro Hiraki\orcidlink{0000-0002-6235-5830}}
\affiliation{Research Institute for Interdisciplinary Science, Okayama University, Okayama 700-8530, Japan}
\author{Yoshitaka Kasamatsu\orcidlink{0000-0003-1477-726X}}
\affiliation{Graduate School of Science, Osaka University, Toyonaka, Osaka 560-0043, Japan}
\author{Shinji Kitao\orcidlink{https://orcid.org/0000-0003-1457-9087}}
\affiliation{Institute for Integrated Radiation and Nuclear Science, Kyoto University,  Osaka 590-0494, Japan}
\author{Adrian Leitner\orcidlink{0009-0007-1156-1881}}
\affiliation{Institute for Atomic and Subatomic Physics, Atominstitut, TU Wien, Vienna 1020, Austria}
\author{Takahiko Masuda\orcidlink{0000-0001-8122-5145}}
\affiliation{Research Institute for Interdisciplinary Science, Okayama University, Okayama 700-8530, Japan}
\author{Nobumoto Nagasawa\orcidlink{0000-0001-5654-7708}}
\affiliation{Japan Synchrotron Radiation Research Insitute,  Kouto,  Hyogo 679-5198, Japan}
\author{Koichi Okai}
\affiliation{Research Institute for Interdisciplinary Science, Okayama University, Okayama 700-8530, Japan}
\author{Ryoichiro Ogake}
\affiliation{Research Institute for Interdisciplinary Science, Okayama University, Okayama 700-8530, Japan}
\author{Martin Pimon\orcidlink{0000-0001-7784-463X}}
\affiliation{Institute for Atomic and Subatomic Physics, Atominstitut, TU Wien, Vienna 1020, Austria}
\author{Martin Pressler\orcidlink{0009-0000-6762-0982}}
\affiliation{Institute for Atomic and Subatomic Physics, Atominstitut, TU Wien, Vienna 1020, Austria}
\author{Noboru Sasao\orcidlink{https://orcid.org/0000-0002-2685-7905}}
\affiliation{Research Institute for Interdisciplinary Science, Okayama University, Okayama 700-8530, Japan}
\author{Fabian Schaden\orcidlink{0000-0001-7154-0440}}
\affiliation{Institute for Atomic and Subatomic Physics, Atominstitut, TU Wien, Vienna 1020, Austria}
\author{Thorsten Schumm\orcidlink{0000-0002-1066-202X}}
\affiliation{Institute for Atomic and Subatomic Physics, Atominstitut, TU Wien, Vienna 1020, Austria}
\author{Makoto Seto}
\affiliation{Institute for Integrated Radiation and Nuclear Science, Kyoto University,  Osaka 590-0494, Japan}
\author{Yudai Shigekawa\orcidlink{0000-0003-2895-4917}}
\affiliation{RIKEN, 2-1 Hirosawa, Wako, Saitama 351-0198, Japan}
\author{Kotaro Shimizu}
\affiliation{Research Institute for Interdisciplinary Science, Okayama University, Okayama 700-8530, Japan}
\author{Tomas Sikorsky\orcidlink{0000-0003-0280-0928}}
\affiliation{Institute for Atomic and Subatomic Physics, Atominstitut, TU Wien, Vienna 1020, Austria}
\author{Kenji Tamasaku\orcidlink{0000-0003-1440-0518}}
\affiliation{RIKEN SPring-8 Center,  Kouto,  Hyogo 679-5148, Japan}
\author{Sayuri Takatori\orcidlink{0000-0002-8705-9624}}
\affiliation{Research Institute for Interdisciplinary Science, Okayama University, Okayama 700-8530, Japan}
\author{Tsukasa Watanabe\orcidlink{https://orcid.org/0000-0002-4986-813X}}
\affiliation{National Institute of Advanced Industrial Science and Technology (AIST),   Ibaraki 305-8563, Japan}
\author{Atsushi Yamaguchi\orcidlink{0000-0002-3981-3872}}
\affiliation{RIKEN, 2-1 Hirosawa, Wako, Saitama 351-0198, Japan}
\author{Yoshitaka Yoda\orcidlink{https://orcid.org/0000-0003-3062-8651}}
\affiliation{Japan Synchrotron Radiation Research Insitute,  Kouto,  Hyogo 679-5198, Japan}
\author{Akihiro Yoshimi\orcidlink{0000-0002-2438-1384}}
\thanks{yoshimi@okayama-u.ac.jp}
\affiliation{Research Institute for Interdisciplinary Science, Okayama University, Okayama 700-8530, Japan}
\author{Koji Yoshimura\orcidlink{0000-0002-2415-718X}}
\affiliation{Research Institute for Interdisciplinary Science, Okayama University, Okayama 700-8530, Japan}


\date{\today}

\begin{abstract}
Recent studies have shown that the lifetime of the $^{229\mathrm{m}}$Th isomer doped in crystals can be shortened by X-ray or laser irradiation, a phenomenon referred to as isomer quenching. We investigate the temperature dependence of X-ray-induced quenching in $^{229}$Th:CaF$_2$ and identify a correlation with the afterglow of the host crystal. These results suggest a mechanism in which X-ray-induced electrons migrate through the lattice and are captured at Th sites, enabling isomer deexcitation via internal conversion through electron--nucleus coupling. This mechanism links nuclear decay to charge carrier dynamics in the host crystal, providing a new interface between nuclear and solid-state physics. The findings offer a pathway to optimize the performance of solid-state nuclear clocks.
\end{abstract}


\maketitle 
\textit{Introduction}. Thorium-229 possesses the energetically lowest first nuclear excited state, an \textit{isomeric state} at approximately $8.356\,\mathrm{eV}$, denoted as $^{\mathrm{229m}}$Th \cite{kraemer2023observation, tiedau2024laser, zhang2024frequency}. It is the only nuclear level that can be excited with the state-of-the-art vacuum ultraviolet (VUV) lasers \cite{zhang2022tunable, thielking2023vacuum, zhu2024measurement, xiao2024proposal, xiao2025continuous}, inspiring the concept of a nuclear clock \cite{peik2003nuclear}. In addition to the $\mathrm{^{229}Th^{3+}}$ ion-trap schemes \cite{peik2003nuclear, campbell2009multiply, campbell2012single, yamaguchi2024laser, scharl2023setup}, the thorium nuclear clock uniquely enables a solid-state implementation by doping $\mathrm{^{229}Th^{4+}}$ into a VUV-transparent crystal \cite{Rellergert2010, kazakov2012performance,  gong2024feasibility}. These complementary approaches have enabled rapid experimental progresses in refining the spectroscopy parameters of $^{\mathrm{229m}}$Th \cite{wense2020229,beeks2021thorium,thirolf2024thorium}.

Recent global collaborative studies  have reduced the $^{\mathrm{229m}}$Th excitation frequency uncertainty from 6\,THz \cite{kraemer2023observation} to 7\,GHz \cite{tiedau2024laser}, 500\,MHz \cite{Elwell:2024qyh}, and 2\,kHz \cite{zhang2024frequency} within less than 4 years, demonstrating significant progresses towards realizing the solid-state nuclear clock. Meanwhile, it is observed that the $^{\mathrm{229m}}$Th lifetime strongly depends on the environment. In the neutral atomic form, it was measured to be $10(1)~\mu\mathrm{s}$ \cite{seiferle2017lifetime}, whereas in an ionic state (within an ion trap), it increases to $2020^{+866}_{-433}$~s \cite{yamaguchi2024laser}. In solid-state environments, the Purcell effect \cite{urbach1998spontaneous} can alter the $^\mathrm{229m}$Th lifetime; in $\mathrm{CaF}_2$, $641(4)$\,s \cite{zhang2024frequency}, and in $\mathrm{LiSrAlF}_6$, $568(24)$\,s \cite{Elwell:2024qyh} have been observed.

In our prior work, we observed a tenfold acceleration of isomer decay in Th:CaF$_2$ crystal under 29.2\,keV X-ray irradiation at room temperature \cite{hiraki2024controlling}. We refer to this phenomenon as isomer quenching, specifically X-ray-induced quenching (XIQ), as it is caused by X-ray irradiation. Similarly, the UCLA group reported a lower-than-expected $^{\mathrm{229m}}$Th yields in Th:LiSrAlF$_6$, conjecturing the existence of quenching channels \cite{Elwell:2024qyh}. Meanwhile, the JILA group observed a short isomer lifetime of $150(11)$\,s in a thin-film $^{229}$ThF$_4$ target, compared to the aforementioned values, suggesting quenching or possible superradiance \cite{zhang2024229thf4}. Additionally, the Leuven group found that the radiative decay fraction of $\mathrm{^{229m}Th}$ varies with host material, influenced by the band gap and defect density \cite{pineda2024radiative}.

These intriguing results stimulated further studies on the quenching mechanism. The UCLA group simulated the crystal’s band structure and proposed an internal conversion (IC) channel, where an intermediate electronic state facilitates isomer de-excitation \cite{morgan2025theory}. They also observed photoinduced quenching in Th:LiSrAlF$_6$ and deduced its quenching cross-section \cite{terhune2025_LIQ}. Meanwhile, the TU Wien group reported laser-induced quenching (LIQ) of $^{\mathrm{229m}}$Th in CaF$_2$, showing that stimulation wavelengths below 420\,nm were effective, while those above 700\,nm were not, suggesting a spectral threshold. They further found that cooling the crystal mitigates LIQ and proposed leveraging it to enhance nuclear clock performance \cite{schaden2025_LIQ}. 
Although significant progress has been made, the microscopic quenching mechanism in solid-state hosts, especially how it varies with temperature, is still not well understood.

In this work, we present a comprehensive experimental study on XIQ of $^{\mathrm{229m}}$Th in CaF$_2$ together with measurements of X-ray-induced afterglow (AG) over a wide temperature range, indicating a close connection between nuclear and electronic degrees of freedom. We propose a mechanism in which the quenching of $^{\mathrm{229m}}$Th occurs via an IC process facilitated by charge capture: electrons excited to conduction band by X-ray irradiation, then thermalized, migrating through the crystal (site-to-site hopping), and eventually trapped at thorium sites, enable IC-mediated relaxation of the isomer. In this regime, the quenching rate $W_q$ for each Th isomer can be expressed as follows:
\begin{equation}
    W_q =n_{\mathrm{e^-}}^{\mathrm{qch}}\cdot\sigma_{\mathrm{qch}}\cdot v(T), \label{equation-1}
\end{equation}
where the $n_{\mathrm{e^-}}^{\mathrm{qch}}$ is the number density of quenching electrons; $\sigma_{\mathrm{qch}}$ is the quenching cross-section; $v(T)$ is the mean diffusion velocity of the thermalized low-energy electrons. The electron-capturing mechanism is analogous to the X-ray-induced charge reduction of Tm$^{4+}$ (Ce$^{3+}$) to Tm$^{3+}$ (Ce$^{2+}$) in CaF$_2$ via electron capture as observed in Refs. \cite{hayes1961EPR, jassemnejad1987photo}. The $v(T)=v_0\exp(-E_a/kT)$ denotes the diffusion velocity of the conduction-band electrons, with $E_a$ representing the activation energy required for migration through the crystal lattice \cite{rodnyi2020physical, lecoq2006inorganic}. Here we focus solely on the temperature dependence of $v$ [See discussion in Quenching model, Eq.~(\ref{equation-5})]. The derivation of the relations connecting $W_q$ to the experimental observables—the $^{\mathrm{229m}}$Th photon yield and lifetime—is also provided in the End Matter.

\begin{figure}[t]
    \centering
    \includegraphics[width=\linewidth]{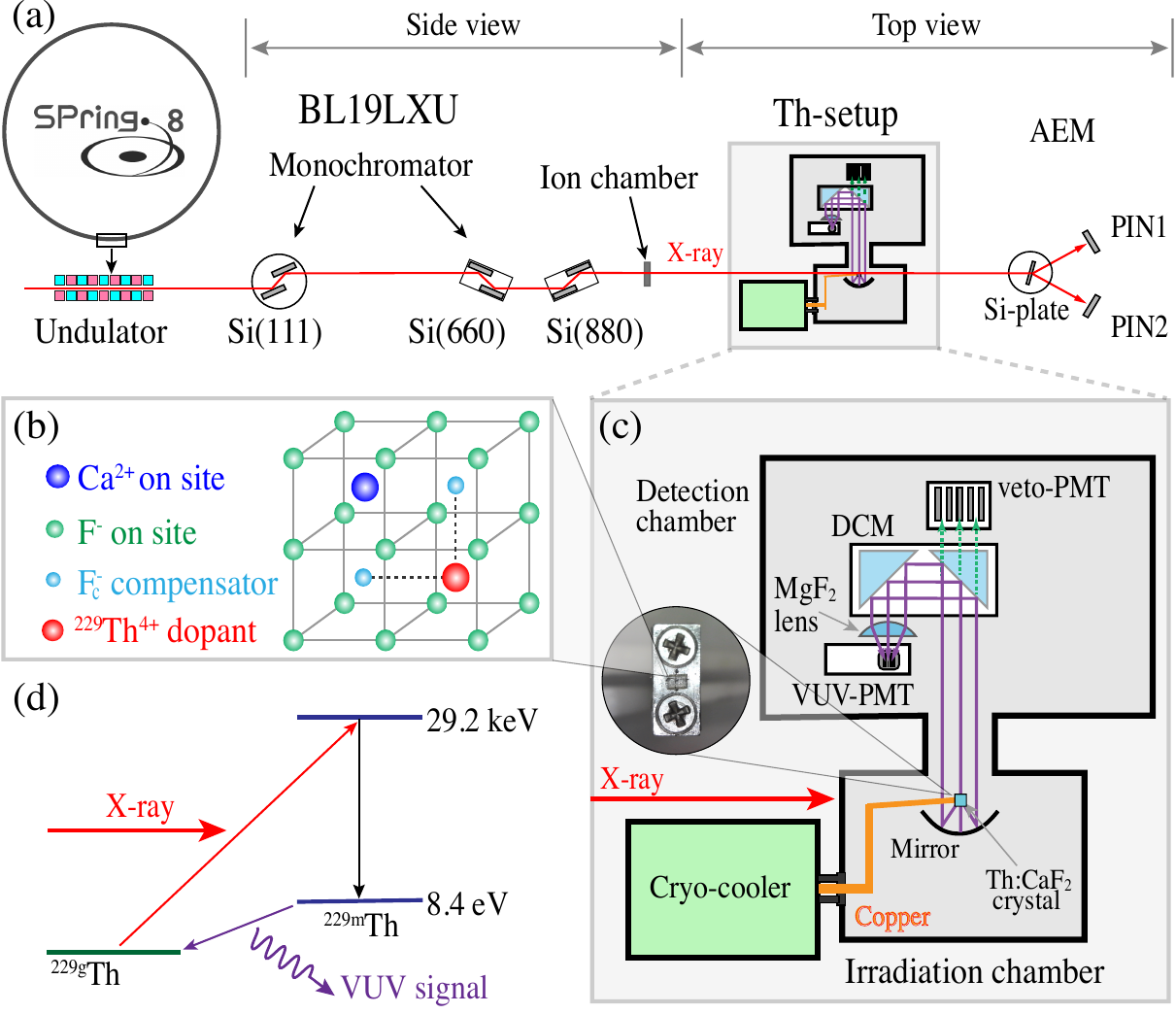}
    \caption{(a) Experimental setup overview. (b) An example of the expected $\mathrm{^{229}Th^{4+}}$ doping configuration in the CaF$_2$ host crystal. (c) Enlarged view of the Th-setup, including a photo of the target crystal. (d) The isomer X-ray pumping scheme via the $29.2\,\mathrm{keV}$  second nuclear excited state.}
    \label{setup}
\end{figure}

\textit{Methods.} Fig.~\ref{setup}\,(a) presents an overview of the experimental setup. The ``Th-setup'' is assembled in the BL19LXU beamline in SPring-8 (a synchrotron radiation facility in Hyogo, Japan). The X-ray beam produced by the 27\,m undulator in the main ring is monochromatized by three pairs of monochromators \cite{yabashi2001design}, resulting in a beam energy width of $31.5$\,meV (FWHM) and a photon flux of $10^{11}$\,ph./s, which were measured by the absolute energy monitoring (AEM) system \cite{masuda2021absolute} and an ionization chamber, respectively. The X-ray beam ($\mathrm{1.2\times0.8\,mm^2}$) is tuned to around $29.2\,\mathrm{keV}$  to populate the $^{\mathrm{229m}}$Th states as indicated in Fig.~\ref{setup}\,(d) \cite{masuda2019nature, hiraki2024controlling}.
 
In this work, we used a $^{229}$Th doped CaF$_2$ crystal (1\,mm$^3$) synthesized by  modified vertical gradient freezing at TU Wien \cite{beeks2023growth}. It was cut from the same ingot ``X2'',  as it was used for the PTB and JILA laser excitation experiments \cite{tiedau2024laser, zhang2024229thf4}. The thorium number density in the crystal can reach $4\cdot 10^{18}$ nuclei per cubic centimeter, and a possible configuration of the the $^{229}$Th$^{4+}$ dopant site in CaF$_2$ host is sketched in Fig.~\ref{setup}\,(b) \cite{kazakov2012performance}.

To detect VUV photons while shielding the detector from X-ray exposure, we implemented a two-chamber vacuum system operating at $\mathcal{O}(10^{-5})\,\mathrm{Pa}$. The “irradiation chamber” is used for the target excitation and isomer population, while the “detection chamber” is for signal detection. A parabolic mirror in the irradiation chamber reflects photons from the target crystal, and custom dichroic mirrors (DCMs) in the detection chamber filter out most background photons. The transmitted photons are focused by an MgF$_2$ lens and detected by a solar-blind VUV-PMT (Hamamatsu R10454). Another UV-sensitive PMT (veto-PMT: Hamamatsu, R11265-203) measures the light bursts in radioluminescence (RL) from the crystal. These two detectors form an anti-coincident measurement scheme to reject the RL background. Moreover, the veto-PMT can measure the afterglow light from the crystal. For the data acquisition, the PMT signals are first amplified, then captured by an oscilloscope (National Instruments, PXIe-5162), and finally saved to PC disk for off-line analysis. 

The cooling of the crystal is achieved by conducting heat from the stainless steel target holder to the cold head (42\,K) of a cryogenic cooler (SunPower, Cryo-GT) via a copper rod, see Fig.~\ref{setup}\,(c). Heating of the crystal is realized by attaching ceramic heater plates to the conducting copper rod. More details about the experimental methods can be found in Supplement Material and Refs. \cite{hiraki2024experimental, guan2025method}.

We applied the nuclear resonance scattering method \cite{masuda2019nature} to populate the $^{\mathrm{229m}}$Th state, as shown in Fig.~\ref{setup}\,(d). In nominal condition, we first irradiate the target crystal for 300\,s with on-resonance X-ray, then block the beam and measure the photons emitted from the crystal for 1800\,s. The X-ray energy is then shifted off-resonance, and the same irradiation-detection cycle is repeated. By applying the background rejection method \cite{hiraki2024experimental} and taking the difference between the on- and off-resonance data, the net isomer signal is extracted from the VUV-PMT counts. The extracted isomer events are binned in 20\,s intervals according to their absolute timestamps, yielding the time-dependent count rate $N_{\mathrm{VUV}}(t)$, see Fig.~\ref{temp_quench}\,(c) for example. In this work, the fitted exponential value of $N_{\mathrm{VUV}}(t)$ at $t=0$ is denoted as $N_0$, which is proportional to the number of produced isomers. 

\begin{figure}[t]
    \centering
    \includegraphics[width=\linewidth]{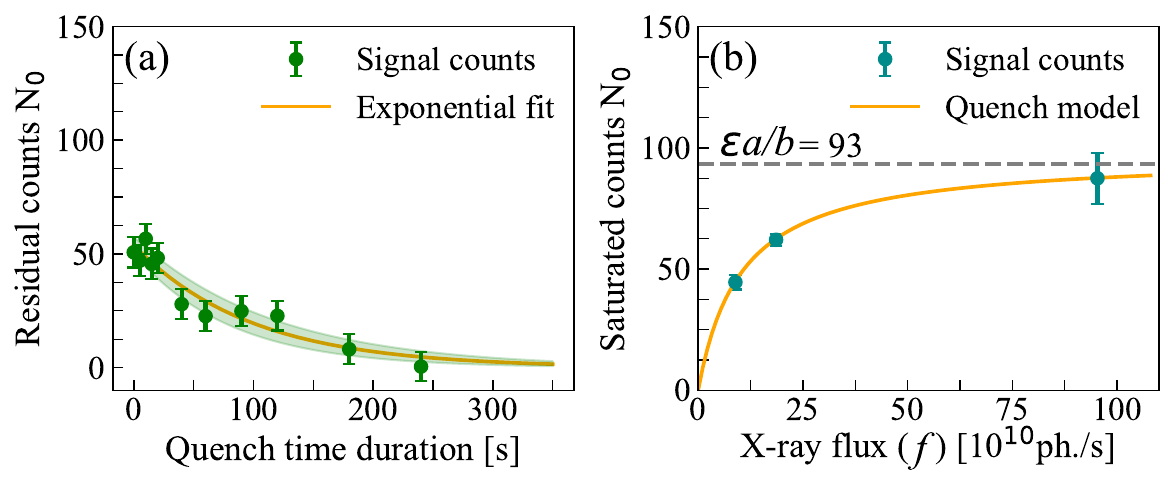}
    \caption{(a) The saturated  $N_0$ along X-ray flux $f$. (b) The residual $N_0$ after quenching X-ray irradiation.}
    \label{X-ray quench}
\end{figure}

\textit{X-ray quenching}.  
When the isomer is pumped by on-resonant X-rays, the isomer-related signal $N_0$ follows the trend in Fig.~\ref{temp_quench}\,(a), eventually reaching a plateau defined as the saturated $N_0$. This saturation reflects a balance between pumping and quenching. The isomer lifetime during irradiation, $\tau_\mathrm{ir}$, can be extracted from this plot. To clearly demonstrate the quenching effect, we performed an additional active quench experiment. The crystal was first irradiated with on-resonance X-rays for 300\,s to saturate the isomer population, then with off-resonance X-rays to actively quench the isomers. The residual isomer signal $N_0$ was measured and plotted in Fig.~\ref{X-ray quench}\,(a) for various quenching durations. Fitting with an exponential decay yields a time constant of $98(25)$\,s, with the shaded area indicating the fitting uncertainty. This corresponds to a shortened isomer lifetime during off-resonant X-ray irradiation, reduced by a factor of 6.5 compared to the radiative lifetime of 641\,s, confirming the XIQ effect.

To investigate how the saturated value of $N_0$ varies with X-ray flux $f$, we measured its dependence, as shown in Fig.~\ref{X-ray quench}\,(b). The observed saturation indicates that the XIQ effect becomes more pronounced with increasing X-ray flux.
We model the flux dependence of the saturated $N_0$ [Fig.~\ref{X-ray quench}\,(b)] as:
\begin{equation}
N_0(f) = \frac{\epsilon\cdot a \cdot f}{W_0 + b \cdot f},\label{equation-2}  
\end{equation}
where $W_0$ is the natural radiative decay-width ($W_0=1/\tau_\mathrm{0}$, and $\tau_0=641(4)$\,s from Ref.~\cite{zhang2024frequency}), and $a$ and $b$ are the coefficients introduced to parametrize the isomer production rate ($a\cdot f$) and quenching rate ($b\cdot f$) along the X-ray flux. Here, we make the detection factor $\epsilon$ (see End Matter) explicit in the equation to clarify the relationship between the number of produced isomers and the observed $N_0$. The derivation of Eq.~\eqref{equation-2} can be found in the End Matter, and its fitting results of Fig.~\ref{X-ray quench}\,(b) are $\epsilon\cdot a=0.014(2)$ and $b=0.00015(4)$ for $f$ in the unit of $10^{10}\mathrm{\,ph./s}$. This fitting result is consistent with the hypothesis expressed by Eq.~\eqref{equation-1} that a higher total power ($f\cdot29.2\,\mathrm{keV}$) deposited in the crystal increases the number density ($n\mathrm{_{e^-}^{qch}}$) of charge carriers, resulting in a higher quenching rate ($W_q$) \cite{lecoq2006inorganic}. 

\begin{figure}[t]
    \centering
    \includegraphics[width=\linewidth]{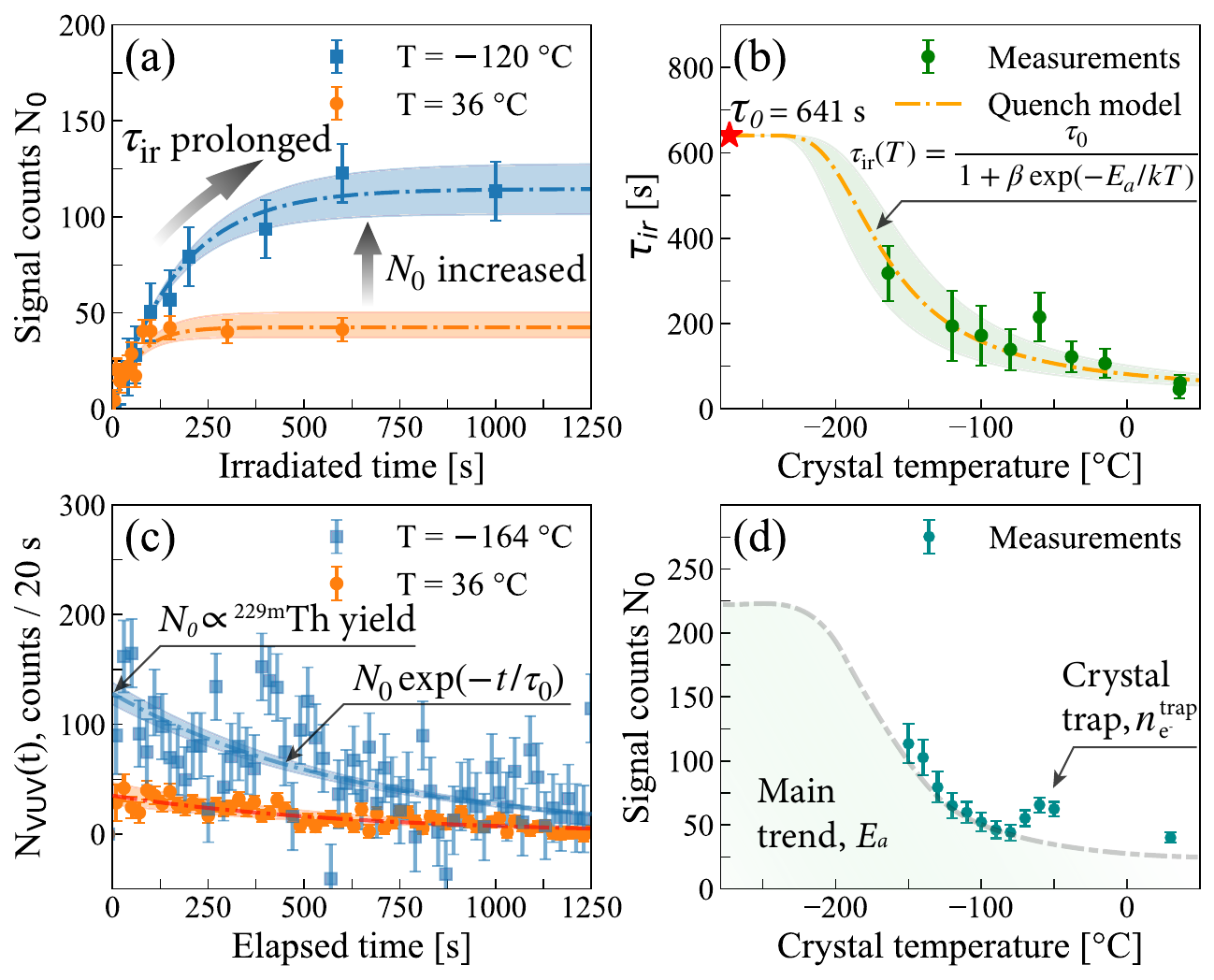}
    \caption{The temperature dependence of the isomer production yield ($N_0$) and the quench lifetime ($\tau_{\mathrm{ir}}$). (a) The isomer production vs. irradiation time. (b) The $\tau_{\mathrm{ir}}$ values measured at different temperatures and fitted with Eq.~\eqref{equation-3}.  (c) Isomer decays as a function of time. (d) The isomer yield measured at different temperature.
    }
    \label{temp_quench}
\end{figure}

\textit{Temperature dependence.} 
To gain further insight into the XIQ process, we measured the isomer yield $N_0$ and lifetime $\tau_{\mathrm{ir}}$ at various crystal temperatures $T$. 
For the data acquisition and analysis, see the Supplemental Material. Fig.~\ref{temp_quench}\,(a) shows the isomer populations measured at $36\,^\circ\mathrm{C}$ and $-120\,^\circ\mathrm{C}$ as a function of X-ray irradiation time. By fitting the population data with  Eq.~\eqref{pumping_expression} \cite{hiraki2024controlling}, we extracted the isomer lifetime $\tau_{\mathrm{ir}}$ for each temperature, as plotted in Fig.~\ref{temp_quench}\,(b). Fig.~\ref{temp_quench}\,(a) and (b) clearly show that cooling the crystal suppresses the XIQ process, as the $N_0$ increased and $\tau_{\mathrm{ir}}$ prolonged at lower temperature. 

The $W_0$ is assumed to be temperature independent \cite{schaden2024laser}; therefore we  deduce that, in the presence of the quenching channel, the isomer lifetime follows $\tau_{\mathrm{ir}}(T) = 1/[W_0 + W_q(T)] = \tau_0 / [1 + W_q(T)/W_0]$. To capture  temperature dependence of $W_q$, we further develop this expression into a thermally activated form (see End Matter):
\begin{equation}
    \tau_{\mathrm{ir}}(T) = \frac{\tau_0}{1+\beta \exp(-E_a/kT)},\label{equation-3}
\end{equation}
where $\beta$ and $E_a$ (activation energy required for site-to-site electron hopping in the crystal) are the fitting parameters and $k=8.617\cdot10^{-5}\,\mathrm{eV/K}$ is the Boltzmann constant. Fitting Eq.~\eqref{equation-3} to the data in Fig.~\ref{temp_quench}\,(b) yields $\beta = 28(11)$ and $E_a = 0.033(6)\,\mathrm{eV}$. The shaded area is the fitting uncertainty of $E_a$, over one standard deviation. The fitting suggests that at lower temperatures, the XIQ can be effectively suppressed such that the $\tau_{\mathrm{ir}}$ approaches $\tau_0$, as indicated by the red star in Fig.~\ref{temp_quench}\,(b).  A notable observation is the unexpected long $\tau_{\mathrm{ir}}$ at $\mathrm{-60\,{^{\circ}C}}$, indicating a reduced quenching rate at this temperature compared to the nearby temperatures.

To examine the unusual value of $\tau_{\mathrm{ir}}$ around $\mathrm{-60\,{^{\circ}C}}$, we measured the isomer yield across a range of temperature values, each time by irradiating the crystal for 900\,s. As exampled in Fig.~\ref{temp_quench}\,(c), the isomer yield $N_0$ was obtained by fixing the lifetime parameter at $\tau_{\mathrm{0}}$ when fitting the isomer decay data. The crystal temperature was scanned from $\mathrm{-50\,{^{\circ}C}}$ to $\mathrm{-150\,{^{\circ}C}}$ in 10 degrees decrements. The resulting data, shown in Fig.~\ref{temp_quench}\,(d), reveals that $N_0$ at $\mathrm{-60\,{^{\circ}C}}$ is indeed larger for neighboring temperatures, while it decreases from $\mathrm{-60\,{^{\circ}C}}$ to $\mathrm{-80\,{^{\circ}C}}$. Below $\mathrm{-80\,{^{\circ}C}}$, the isomer yield tends to increase with lowering temperature. The trend observed in $N_0(T)$ is consistent with the observed temperature-dependent variation in $\tau_{\mathrm{ir}}(T)$. 

\begin{figure}[t]
    \centering
    \includegraphics[width=\linewidth]{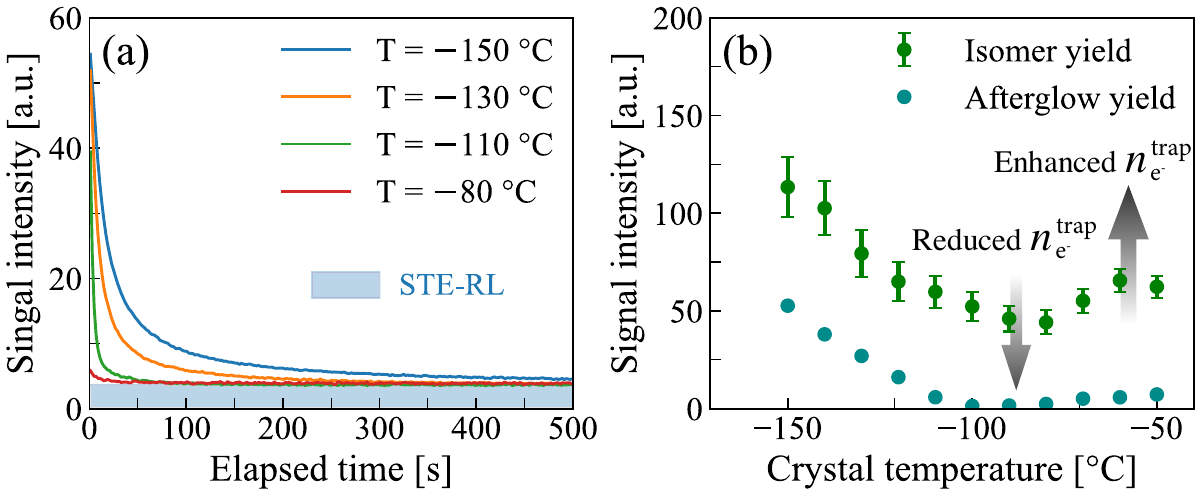}
    \caption{Afterglow yield from the $^{229}$Th:CaF$_2$ crystal after X-ray irradiation. (a) The afterglow of the X2 crystal following X-ray irradiation for 900\,s at $-80$, $-110$, $-130$, and $-150$\,$\mathrm{^\circ  C}$ with the STE-RL offset (see text). (b) The scaled integrated afterglow yield plotted together with the isomer yield from Fig.~\ref{temp_quench}\,(d).
    }
    \label{afterglow}
\end{figure}

\textit{Afterglow and quenching. }
The veto-PMT in Fig.~\ref{setup}\,(c) detects ultraviolet fluorescence emitted from the crystal after X-ray irradiation, referred to as afterglow (AG). AG is a type of luminescence from CaF$_2$ \cite{rao1971afterglow}, originating from delayed recombination of trapped electrons and holes. The veto-PMT also detects self-trapped exciton (STE) luminescence in radioluminescence (RL) as a time-independent offset, denoted as STE-RL.  Fig.~\ref{afterglow}\,(a) shows four examples of AG from the X2 crystal after 900\,s of X-ray irradiation, illustrating increased yield from $\mathrm{-80\,{^{\circ}C}}$ to $\mathrm{-150\,{^{\circ}C}}$. After removing the STE-RL components, the summed net afterglow yields are plotted as cyan points in Fig.~\ref{afterglow}\,(b), alongside isomer yield (green points) at various temperatures.  The similar trends in isomer yield and AG indicate a link between isomer quenching and electron-hole dynamics.

\begin{figure}[t]
    \centering
    \includegraphics[width=\linewidth]{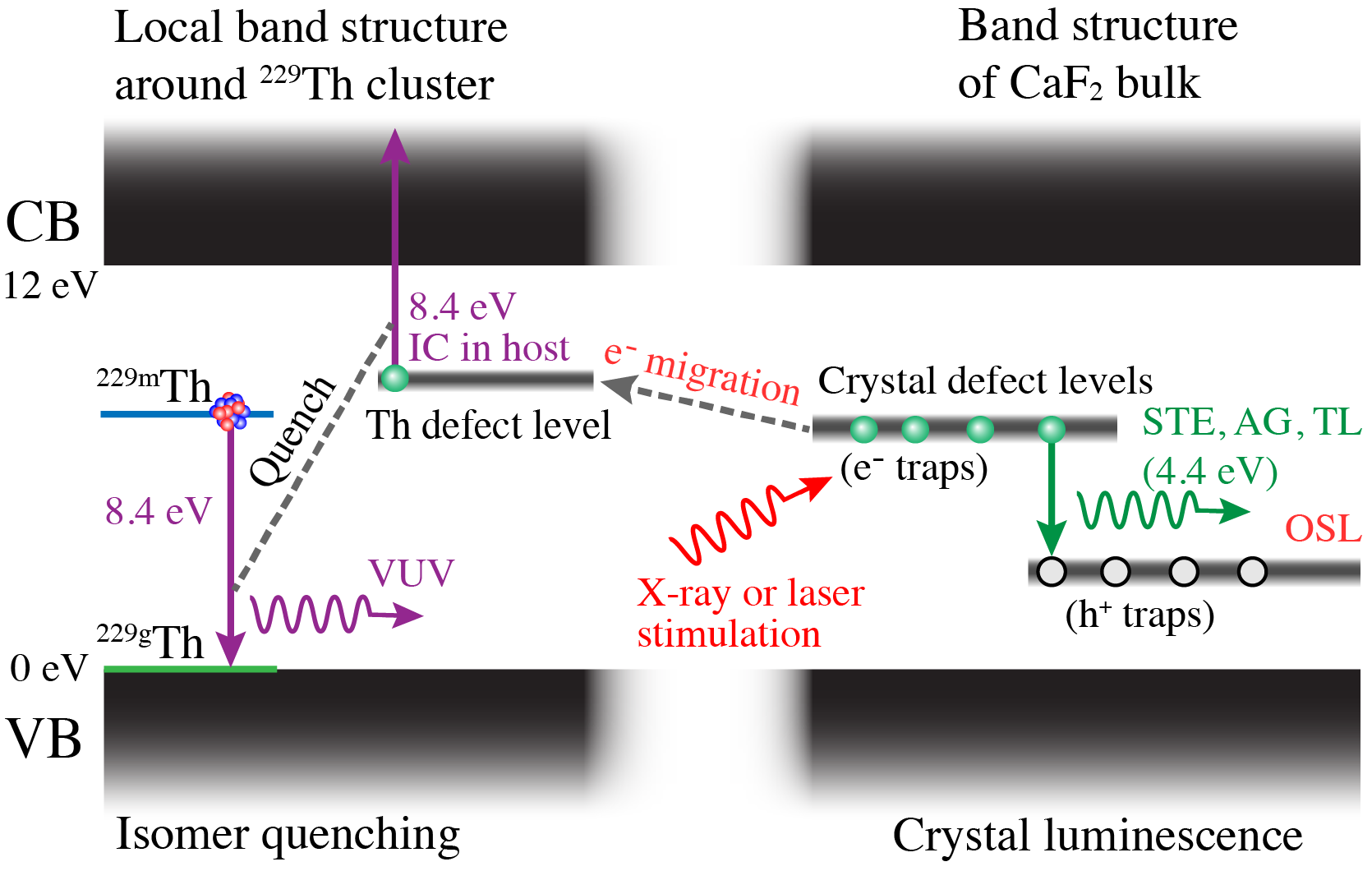}
    \caption{Conceptual band structure of Th:CaF$_2$. CB: conduction band; VB: valence band. The right side illustrates the bulk CaF$_2$ band structure, where STE, AG, and TL arise from recombination of trapped charge carriers. The left side shows the localized band structure near the Th dopant. The $^{\mathrm{229m}}$Th can decay via VUV emission or by transferring energy to a trapped electron at an intermediate level. External stimulation can activate trapped electrons ($\mathrm{e^-}$), enabling migration. The migrating electron may recombine with a hole ($\mathrm{h^+}$), producing OSL, or be captured by a thorium defect, thereby facilitating isomer quenching.
    }
    \label{band_model}
\end{figure}

To accommodate this similar temperature trend, we introduce a temperature-dependent correction to the $n\mathrm{^{qch}_{e^-}}$ in the quenching model Eq.~\eqref{equation-1}. Accordingly, the temperature dependence of $N_0(T)$ is formulated as
\begin{equation}
N_0(T) = \frac{\epsilon\cdot R}{W_0 + [n\mathrm{_{e^{-}}} - n{\mathrm{_{e^{-}}^{trap}}}(T)]\cdot \sigma_{\mathrm{qch}}\cdot v(T)}, \label{equation-4}
\end{equation}
where the $R$ is the isomer population rate, the $n\mathrm{_{e^-}}$ is the number density of migrating electrons which only depends on X-ray power in this model,  $n\mathrm{_{e^-}^{trap}}$ is the number density of electrons trapped in crystal defects. The afterglow yield is proportional to the value of $n\mathrm{_{e^-}^{trap}}$. In this expression, $[n\mathrm{_{e^{-}}} - n{\mathrm{_{e^{-}}^{trap}}}(T)]$ stands for the $n\mathrm{_{e^-}^{qch}}$ in Eq.~\eqref{equation-1}. 
At $-80\,\mathrm{^\circ C}$, the small afterglow yield implies a low $n\mathrm{_{e^-}^{trap}}$, which increases the denominator $[n\mathrm{_{e^{-}}} - n{\mathrm{_{e^{-}}^{trap}}}(T)]$ in Eq.~\eqref{equation-4}, resulting in a reduced $N_0(T)$. A similar explanation applies to the $N_0(T)$ peak at $-60\,\mathrm{^\circ C}$: increased $n\mathrm{_{e^-}^{trap}}$ lowers $n\mathrm{_{e^-}^{qch}}$ and $W_q$, enhancing both AG and $N_0$. Thus, isomer lifetime, yield, and AG intensity exhibit similar temperature trends, as shown in Fig~\ref{temp_quench}\,(d) and \ref{afterglow}\,(b).

\textit{Quenching model.}
To understand the results heuristically introduced above, we propose a quenching process within the host crystal band gap, conceptually depicted in Fig.~\ref{band_model}, where the VB stands for the crystal valence band and the CB for the conduction band. The right side of the figure shows the band structure of bulk CaF$_2$ (without $\mathrm{^{229}Th^{4+}}$ doping), highlighting crystal defect levels within the band gap which trap electrons and holes. The recombination of these trapped charge carriers leads to various luminescence phenomena, including RL, AG. The left side of the figure illustrates the localized band gap structure around the thorium dopant (see Fig.~\ref{setup}\,(b) or Ref. \cite{nalikowski2025embedded}), where the isomeric nuclear level ($8.4\,\mathrm{eV}$) and an intermediate electronic level reside within the band gap \cite{morgan2025theory}. 

The $\mathrm{^{229}Th^{4+}}$ defect level shown in Fig.~\ref{band_model} may originate from local charge compensation, estimated near $10\,\mathrm{eV}$ in Refs. \cite{dessovic2014229thorium, nickerson2020nuclear, nickerson2021driven} and around $7.5\,\mathrm{eV}$ in Ref. \cite{morgan2025theory}, or from charging of the thorium defect itself. The isomer can decay via VUV photon emission or through energy transfer to a trapped electron at an intermediate state. In the latter case, a migrating electron captured at the $^{229}$Th site enables quenching via spatial wavefunction overlap. The nucleus transfers energy to the electron, promoting it to the conduction band—analogous to internal conversion in atoms \cite{wense2016nature}, but occurring within the solid-state host \cite{morgan2025theory}. This model also accounts for the $\sim 2\,\mathrm{eV}$ spectral threshold observed in LIQ \cite{schaden2025_LIQ}, attributed to the energy needed to release crystal trapped electrons, which should be comparable with the threshold for optically stimulated luminescence (OSL) in CaF$_2$ \cite{yukihara2011optically, nanto2015optically}.

As a recap of the XIQ experiments and the proposed quenching model, the quenching rate—complete form of Eq.~\eqref{equation-1}—is now given by
\begin{equation}
    W_q = [n\mathrm{_{e^{-}}} - n{\mathrm{_{e^{-}}^{trap}}}(T)]\cdot \sigma_{\mathrm{qch}}\cdot v_0\exp(-E_a/kT).  \label{equation-5}
\end{equation}
Here, $[n\mathrm{_{e^{-}}} - n{\mathrm{_{e^{-}}^{trap}}}(T)]$ denotes the number density of electrons available for quenching, and the Arrhenius factor $\exp(-E_a/kT)$ describes the temperature-dependent electron migration. The latter determines the main temperature trend of $^{229\mathrm{m}}$Th quenching, while the former modulates local temperature deviations from this trend. The temperature trend of the isomer yield in Fig.~\ref{temp_quench}\,(d) can be explained by these factors.
A more detailed analysis of potential X-ray flux dependence, including carrier-density effects and internal field formation, is beyond the scope of this study and will be addressed in future work.

\textit{Summary.} In this Letter, we present a comprehensive study of X-ray-induced quenching (XIQ) of $^{\mathrm{229m}}$Th in a CaF$_2$ host, revealing a strong coupling between electronic and nuclear degrees of freedom. We find that higher X-ray flux and elevated temperatures enhance isomer quenching. The temperature dependence of the quench lifetime, isomer yield, and crystal afterglow follows similar trends, motivating a quenching model in which electrons are captured at thorium sites to facilitate quenching. Fitting the lifetime data within this framework yields an activation energy of $E_a = 0.033(6)\,\mathrm{eV}$, corresponding to a characteristic temperature of $383(70)\,\mathrm{K}$. The Arrhenius factor associated with this activation energy dominates the overall temperature dependence of XIQ, with deviations attributed to temperature-dependent electron trapping by crystal. The reported mechanism may provide a guide for future solid-state nuclear clock design and offers a probe of nuclear–host coupling.

\begin{acknowledgements}
    {The author MG thanks Dr. Wang Jing for the constructive discussions. This work was supported by JSPS KAKENHI Grant Numbers JP21H04473, JP23K13125, JP24K00646, JP24H00228, JP24KJ0168. This work was also supported by JSPS Bilateral Joint Research Projects No. 120222003. This work has been funded by the European Research Council (ERC) under the European Union’s Horizon 2020 research and innovation programme (Grant Agreement No. 856415) and the Austrian Science Fund (FWF) [Grant DOI: 10.55776/F1004, 10.55776/J4834, 10.55776/ PIN9526523]. }    
\end{acknowledgements}

\bibliographystyle{apsrev4-2}
\bibliography{XIQ_229mTh}

\clearpage

\section{End Matter}
\section{A. Theoretical preparation}
\subsection{A.1 The dynamic of isomer states, \(N_{\mathrm{iso}}(t)\)}

The dynamic changes of the amount of $^{\mathrm{229m}}$Th in the crystal host  $N_{\mathrm{iso}}(t)$ can be expressed in a differential equation
\begin{equation}
    \dfrac{dN_{\mathrm{iso}}(t)}{dt} =  \xi R  - W_0N_{\mathrm{iso}}(t) - \eta W_qN_{\mathrm{iso}}(t),\hspace{0.5em} ( \xi,\,\eta=0\,\mathrm{or}\,1) \label{isomer_equation}
\end{equation}
where the $R$ is the isomer production rate when the X-ray beam is in resonance with the thorium nuclei, $W_0$ is the isomer radiative decay rate via emitting a VUV photon, and $W_q$ is the isomer decay rate via the additional quenching channel. Depending on the resonance and exposure of the X-ray beam, the boolean values $ \xi$ and $\eta$ are switched between 1 and 0. 

When the X-ray beam is on-resonance, the pumping and quenching channel exist simultaneously ($ \xi=\eta=1$), which leads to the isomer population dynamics depicted in Fig.~\ref{temp_quench}\,(a), and which reads as
\begin{equation}
    N_{\mathrm{iso}}(t) = \frac{R}{W_0+W_q} [1-\exp(-(W_0+W_q)t)]. \label{pumping_expression}
\end{equation}
Therefore, the isomer lifetime during X-ray irradiation $\tau_{\mathrm{ir}}$ satisfies $\tau_{\mathrm{ir}}=\frac{1}{W_0+W_q}$, and after long time irradiation, the saturated isomer yield is $N_0=\frac{R}{W_0+W_q}$.

If the X-ray energy is shifted to off-resonant, the isomer pumping stopped ($ \xi=0$) while isomer quenching still occurs ($\eta=1$), then the solution of Eq.~\eqref{isomer_equation} reads as
\begin{equation}
    N_{\mathrm{iso}}(t) = N_0 \exp[ -(W_0+W_q)t ]. 
\end{equation}
This isomer quench process is shown in Fig.~\ref{X-ray quench}\,(a).

When the X-ray is turned off, both $\xi$ and $\eta$ related terms disappear in Eq.~\eqref{isomer_equation} and we obtain the spontaneous decay of the isomer signal
\begin{equation}
    N_{\mathrm{iso}}(t) = N_0 \exp(-W_0t).
\end{equation}
This corresponds to the spontaneous radiative decay of the isomeric states, as exemplified in Fig.~\ref{temp_quench}\,(c).

\subsection{A.2 Isomer pumping rate, \(R\)}

The isomer production rate $R$ via the NRS method had been introduced in our previous works, including in $\mathrm{Th(NO_3)_4}$ precipitate \cite{masuda2019nature}, and in $^{229}$Th:CaF$_2$  crystal \cite{hiraki2024controlling}. The isomer production-related nuclear energy levels are shown in Fig.~\ref{setup}\,(d), and the rate is the product of the pumping rate of the $29.2\,\mathrm{keV}$  level $R_{\mathrm{NRS}}$ and the in-band transition branching ratio $Br\mathrm{_{tot}^{in}}$. Considering the attenuation of the X-ray intensity inside the target by absorption length $l_{_X}$, and denoting the X-ray flux
 as $f$, the  isomer production rate is
\begin{align}
        R = Br_{\mathrm{tot}}^{\mathrm{in}} n_{\mathrm{Th}} \sigma_{\mathrm{eff}} f l_{_X} (1-\mathrm{e}^{-L/l_{_X}}), \label{isomer_rate}
\end{align}
where the $\sigma_{\mathrm{eff}}$ is called the effective cross-section, expressed as
\begin{equation}
\sigma_{\mathrm{eff}}=\frac{\lambda^2_{\mathrm{2nd}}}{4}\frac{\Gamma_\gamma^{\mathrm{cr}}}{\sqrt{2\pi}\sigma_{\text{X-ray}}}. \label{effective_cross_section}
\end{equation}
The quantities involved in this calculation are listed in Tab.~\ref{variables}. For the flux $f$ dependence, we denote Eq.~\eqref{isomer_rate} as $R=a\cdot f$ and using the values in Tab.~\ref{variables}, we get  $R\simeq 8,706\,\mathrm{isomer/s}$ and $a = 5.21\cdot10^{-8}$. Here, it should be noted that the fitted parameter in Fig.~\ref{X-ray quench}\,(b) is determined with X-ray flux in unit of $10^{10}$\,ph./s, the calculated value of $a$ should be converted as $a = 521$. By applying the detection factor $\epsilon = 1.6 \times 10^{-4}$ (see Supplemental material), the calculated $\epsilon \cdot a$ becomes 0.0808. The factor of 6 discrepancy between the calculated 0.0808 and fitted value of $\epsilon\cdot a=0.014(2)$ may arise from run-dependent factors such as device alignment, background rejection, and X-ray flux variations.

In Fig.~\ref{X-ray quench}\,(b), the first two data points were measured with an X-ray energy width of $\sigma_{\text{X-ray}}=13.4\,\mathrm{meV}$, while the third was measured at $\sigma_{\text{X-ray}}=40.0\,\mathrm{meV}$. The isomer yield correction for the third point  is applied by compensating the difference of its X-ray energy width $\sigma_{\text{X-ray}}$ using Eq.~\eqref{effective_cross_section}, by multiplying the third data point with a factor of $\mathrm{40.0\,meV/13.4\,meV}$.

\begin{table}[t]
    \centering
   \caption{The variables to calculate the isomer production rate.}\label{variables}   
    \begin{tabular}{ccc}
    \hline
    \hline
       Symbol  &  Value & Note \\
        \hline
       $\Gamma^{\mathrm{cr}}_\gamma$  & 1.70\,neV & cross bandwidth \\
       $\lambda_{\mathrm{2nd}}$ & 42.48\,pm & X-ray wavelength\\
       $Br\mathrm{_{tot}^{in}}$ & 0.72 & in-band branch ratio \\
       $\sigma_{\text{X-ray}}$ & $13.4\,\mathrm{meV}$  & X-ray energy width\\
       $f$ & $16.7\cdot 10^{10}$\,ph./s & X-ray flux \\
       $n\mathrm{_{Th}(X2)}$ & $4.0\cdot10^{18}$\,cm$^{-3}$ & Th$^{4+}$ number density \\
       $L(\mathrm{X2})$ & 1.2\,mm & X2 length\\
       $l_{_X}$($29.19\,\mathrm{keV}$) & 1.34\,mm & X-ray attenuation length\\
        \hline
        \hline
    \end{tabular}
\end{table}

\subsection{A.3 Isomer VUV decay rate, \(W_0\)}
The radiative decay rate of $^{\mathrm{229m}}$Th in CaF$_2$ crystal environment (include the Purcell effect \cite{urbach1998spontaneous}) had been experimentally measured in the SPring-8 \cite{hiraki2024controlling}, PTB \cite{tiedau2024laser, schaden2025_LIQ}, JILA \cite{zhang2024frequency} experiments. The lifetime values are consistent within measurement uncertainty. Among them, the most precise value is $\tau_0 =641(4)$\,s, corresponding to a radiative decay rate of $W_0= 1/\tau_0 =1/641$\,Hz, which is adopted in this report. 

\subsection{A.4 Isomer quenching rate, \(W_q\)}
In the proposed model, Fig.~\ref{band_model}, the isomer quenching happens when an electron is captured by the intermediate defect level in the thorium cluster. The conduction band electrons promoted by X-ray beam would either trapped by the crystal defect levels or fall into the Th intermediate level. Denote the total number density of movable electron as $n_{\mathrm{e^-}}$ \cite{lecoq2006inorganic}, and the number density of electrons trapped by CaF$_2$ crystal defect level as $n{\mathrm{^{trap}_{e^-}}}$, the number density of electron available to be captured by the Th intermediate defect level is $n\mathrm{_{e^-}^{qch}}$. These values satisfy the relation
\begin{equation}
    n\mathrm{_{e^-}}= n\mathrm{_{e^-}^{trap}} + n\mathrm{_{e^-}^{qch}} \label{sum_Ne}
\end{equation}

For a single $^{\mathrm{229m}}$Th nucleus, the quench rate could be expressed as 
\begin{align}
W_q(f,T) &= n\mathrm{_{e^-}^{qch}}(f,T)\cdot \sigma_{\mathrm{qch}}\cdot v(T),\label{single_quench_rate} 
\end{align}
where the $\sigma\mathrm{_{qch}}$ is the quenching cross-section, and $v(T)$ is the mean thermal diffusion velocity of the electron in the crystal. This relation reveals that the quench rate is determined by the number density and velocity of the electrons, which are both highly temperature-sensitive. 

The term $v(T)$ in Eq.~\eqref{single_quench_rate}  resembles a mobility term; however, it does not represent conventional electron mobility defined by drift under an external electric field. Instead, in CaF$_2$—a wide-band-gap insulator with strong electron–phonon coupling \cite{hayes2012defects}—electron transport is better described by thermally activated hopping between localized sites which arise from electron-lattice interaction. This process is modeled using an Arrhenius-type hopping probability, resulting in a diffusive mean velocity described by 
\begin{equation}
    v(T) =  v_0\exp(-E_a/kT), \label{electron_migration}
\end{equation}
where $v_0$ is the velocity pre-factor, $E_a$ is the activation energy required for hopping, $k = 8.617 \cdot 10^{-5}\,\mathrm{eV/K}$ is the Boltzmann constant, and $T$ is the crystal temperature \cite{rodnyi2020physical}. 

\section{B. Experimental Observables}

\subsection{B.1 Isomer yield along X-ray flux, \(N_0(f)\)}

For the \textit{X-ray quenching} experiment at a constant temperature, the $v(T)$ is constant. Therefore, the quenching is determined by the number density of electrons participating in the quenching reaction, thus  $ W_q\propto n\mathrm{_{e^{-}}^{qch}}(f)\propto f$. Under X-ray irradiation, increased power promotes a higher rate of electron excitation into the conduction band, leading to an enhanced quenching rate. Therefore, the coefficient $b$ in  Eq.~\eqref{equation-2} is the equivalent to the right side of Eq.~\eqref{single_quench_rate} without the implicitly contained $f$.

After sufficient irradiation time, the saturated isomer yield is determined by the production rate $R$ and the total decay rate of the isomer state, $W_0 + W_q$. Considering Eq.~\eqref{pumping_expression} and the detection factor $\epsilon$, the $N_0(f)$ can be expressed as
\begin{equation}
    N_0(f) = \frac{\epsilon\cdot R(f)}{W_0+W_q(f)} = \frac{\epsilon\cdot a\cdot f}{W_0 + b\cdot f},
\end{equation}
where $a$ and $b$ correspond to the production  and quenching rates of isomers per unit X-ray flux. The value of $a$ is introduced in A.2, whereas $b$ is currently under study.

\subsection{B.2 Quenching lifetime along temperature, \(\tau\mathrm{_{ir}}(T)\)  }

In the \textit{temperature dependence} measurements, the X-ray flux is fixed, then the quenching rate is mainly determined by the migration of the electrons. The quench lifetime is 
\begin{equation}
    \tau_{\mathrm{ir}} = \frac{1}{W_0+W_q} = \frac{1}{W_0}\frac{1}{1+W_q/W_0}. \label{split_W}
\end{equation} 
Given the $W_q$ defined in Eq.~\eqref{single_quench_rate}-\eqref{electron_migration}, the above expression can be written as
\begin{equation}
    \tau_{\mathrm{ir}} (T)=  \frac{\tau_0}{1+\beta \exp(-E_a/kT)}, \label{temp_model}
\end{equation}
where $\tau_0=1/W_0$ and $\beta=n\mathrm{_{e^-}^{qch}}(T)\sigma_{\mathrm{qch}} v_0 /W_0$. Here, \(\beta\) is actually temperature-dependent variable due to \(n\mathrm{_{e^-}^{qch}}(T)\). However, to capture the main temperature trend, we assume it to be a constant parameter. Therefore, we could see that the quench lifetime during irradiation becomes shorter at higher temperatures, due to the exponential factor in the denominator, as shown in Fig.~\ref{temp_quench}\,(b). The Fitting result of the Eq.~\eqref{temp_model} yields that $\beta=28(11)$, and $E_a=0.033(6)\,\mathrm{eV}$.

\subsection{B.3 Isomer yield along temperature, \(N_0(T)\)}

During irradiation, X-ray creates electron-hole pairs. Most of them will recombine swiftly, thus the crystal emits photoluminescence, while the separated electrons and holes will migrate inside the crystal. The migrating electron will either be trapped by the crystal defects or captured by the intermediate level which facilitates  isomer quenching, and their number density are denoted as $n{\mathrm{_{e^-}^{trap}}}$ and $n{\mathrm{_{e^-}^{qch}}}$, respectively. These electron number density satisfy the relation of Eq.~\eqref{sum_Ne}. Therefore, the quenching rate for an isomeric nucleus becomes
\begin{equation}
    W_q = [n{\mathrm{_{e^-}^{}}}-n{\mathrm{_{e^-}^{trap}}}(T)]\cdot \sigma_{\mathrm{qch}}\cdot
    v(T),
\end{equation}
and the saturate isomer yield after long time irradiation is
\begin{equation}
    N_0(T) = \frac{\epsilon\cdot R}{W_0 + [n{\mathrm{_{e^-}^{}}} - n{\mathrm{_{e^-}^{trap}}}(T)]\cdot \sigma_{\mathrm{qch}}\cdot
    v(T)}.
\end{equation}

\onecolumngrid
\begin{center}
{\large{\textbf{Supplement Material for:\\
\vspace{0.5em}
``X-ray-induced quenching of the $^{229}$Th clock isomer in CaF$_2$''}}}

\vspace{1em}
Ming Guan$^1$,\\
\textit{${}^1$on behalf of 30 collaborators from 8 institutes listed in the main manuscript}\\
(Date: August 8, 2025)

\end{center}
\twocolumngrid


The main manuscript presents the key experimental results of the X-ray-induced quenching (XIQ) of the $^{229}$Th isomer. This Supplemental Material provides the essential technical details, including (A) background reduction and data analysis, (B) data acquisition procedures, and (C) cryogenic cooling methods. To avoid confusion, references enclosed in quotation marks, such as ``Fig.'', refer to figures in the main manuscript.

\section{A. Background and signal processing}

\begin{figure}[!t]
    \centering
    \includegraphics[width=\linewidth]{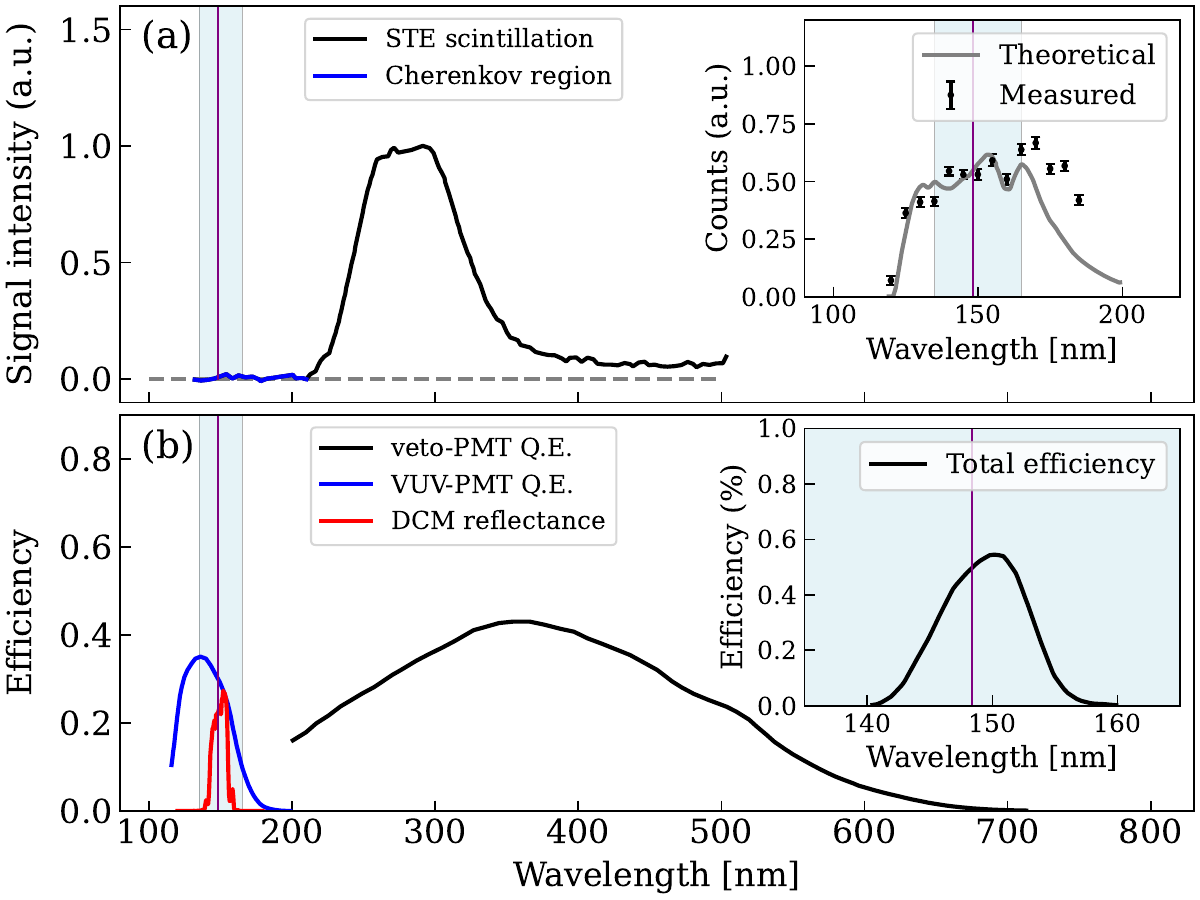}
    \caption{The violet vertical line near 148.4\,nm represents the isomer signal \cite{Elwell:2024qyh}. The shade area is the suggested signal region in Ref.~\cite{Seiferle2019}.  (a) The radioluminescence spectrum of the $^{229}$Th:CaF$_2$ crystal is measured by TU Wien group \cite{stellmer2015radioluminescence}, where the distinct STE band was observed but the signal counts in Cherenkov region is barely discernible from zero. The inset plot shows the updated measurement and calculation result for the Cherenkov region \cite{beeks2023growth}.
    The (a.\,u.) on both vertical axes stands for arbitrary units. (b) Efficiencies of optical devices used in this measurement. Q.\,E.: quantum efficiency. Q.\,E. data were taken from Hamamatsu Photonics \cite{hamamatsu2007photomultiplier}. DCM reflectance is the total reflectance of 4 dichroic mirrors, measured by our group \cite{okai2025_thesis}. The inset of (b) shows the total detection efficiency of the setup near the isomer signal wavelength region, exhibiting a 75\% reduction compared to the efficiency reported in Ref.~\cite{hiraki2024controlling}.
}
    \label{fig:spectra_efficiency}
\end{figure}

In this work, we employed two methods to reduce background events during the measurement of the VUV signals: the wavelength-selective reflection of dichroic mirrors (DCMs) and a two-detector anti-coincidence technique. As illustrated in Fig.~\ref{fig:spectra_efficiency}, the isomer signal is indicated by the vertical violet line at 148.4\,nm \cite{Elwell:2024qyh}. The shaded area represents the $\pm5\sigma$ uncertainty range of the isomer energy, as determined by energy spectroscopy of the internal conversion electrons reported by in Ref.~\cite{Seiferle2019}. All experimental measures were designed to enhance the detection efficiency for the isomer signal wavelength and to effectively reject background-triggered events.

During the isomer signal measurement, two primary types of background are present. The first is radioluminescence, which primarily originates from interactions between the $\alpha$ and $\beta$ particles emitted from the $^{229}$Th decay chain and the CaF$_2$ crystal matrix \cite{stellmer2015radioluminescence}. The scaled  radioluminescence spectrum of $^{229}$Th:CaF$_2$ is shown in Fig.~\ref{fig:spectra_efficiency}\,(a). As illustrated, the  radioluminescence consists of two components: Cherenkov radiation and self-trapped exciton (STE) scintillation, each dominating in different wavelength regions \cite{beeks2023growth}. The STE scintillation peaks at 280\,nm, while the Cherenkov radiation spectrum lies within the 120--200\,nm range, see the inset of Fig.~\ref{fig:spectra_efficiency}\,(a) \cite{Beeks:2022hkjthesis, beeks2023growth}. The second background source is X-ray-induced fluorescence, commonly referred to as afterglow, which decays to negligible levels within several minutes after the X-ray beam is turned off.

\subsection{A.1 The background suppression}

During the signal measurement, the goal was to maximize the detection of isomer VUV photons while minimizing background photon contributions. As a first step, we implemented an assembly of four DCMs in the detection setup, each providing relatively high reflectance near the isomer signal wavelength region. An example of the cumulative four-fold reflectance from the DCM assembly is shown as the red curve in Fig.~\ref{fig:spectra_efficiency}\,(b).

Secondly, we employed a solar-blind photomultiplier tube (VUV-PMT, Hamamatsu R10454) to detect photons in the wavelength range near that of the isomer VUV emission. The quantum efficiency (Q.\,E.) of the VUV-PMT is shown as the blue curve in Fig.~\ref{fig:spectra_efficiency}\,(b), exhibiting an efficiency of approximately 30\,\% around the isomer signal wavelength and rapidly decreasing to near zero for longer wavelengths ($\sim180\,\mathrm{nm}$).

The wavelength-dependent total detection efficiency, shown in the inset of Fig.~\ref{fig:spectra_efficiency}\,(b), is primarily determined by the combined wavelength dependence of the DCM assembly reflectance and the VUV-PMT quantum efficiency. As shown in the inset, the total detection efficiency exhibits a peak value of approximately $\epsilon_\mathrm{det.}=$0.5\,\% with an effective full width at half maximum of about 10\,nm for VUV signal photons.

Comparing this efficiency curve with the spectral distribution of the Cherenkov radiation background, it is clear that a significant number of Cherenkov photons still fall within the detection window and can overwhelm the isomer VUV signal photons.

In the main manuscript and the End Matter, we define the detection efficiency $\epsilon$ as the factor that relates the number of produced isomers to the observed value $N_0$ at the initial time $t = 0$, after X-ray pumping is switched off [see Eq.~(2) in the main text]. In this scheme, $\epsilon$ is composed of two components: the intrinsic detection efficiency $\epsilon_\mathrm{det.}$ and the time-binning factor $\epsilon_\mathrm{bin}$. The time-binning factor corresponds to the fraction of the isomer signal contained within the first 20-second time bin relative to the total signal over the full decay curve, and is given by $[1 - \exp(-20/\tau)] = 0.031$, when we use $\tau=641$\,s. The value of $\epsilon$ is then estimated as $\epsilon = \epsilon_\mathrm{det.} \cdot \epsilon_\mathrm{bin} = 1.6 \times 10^{-4}$, as used in the End Matter.

\subsection{A.2 The background rejection}
\begin{figure}[t]
\centering
\includegraphics[width=\linewidth]{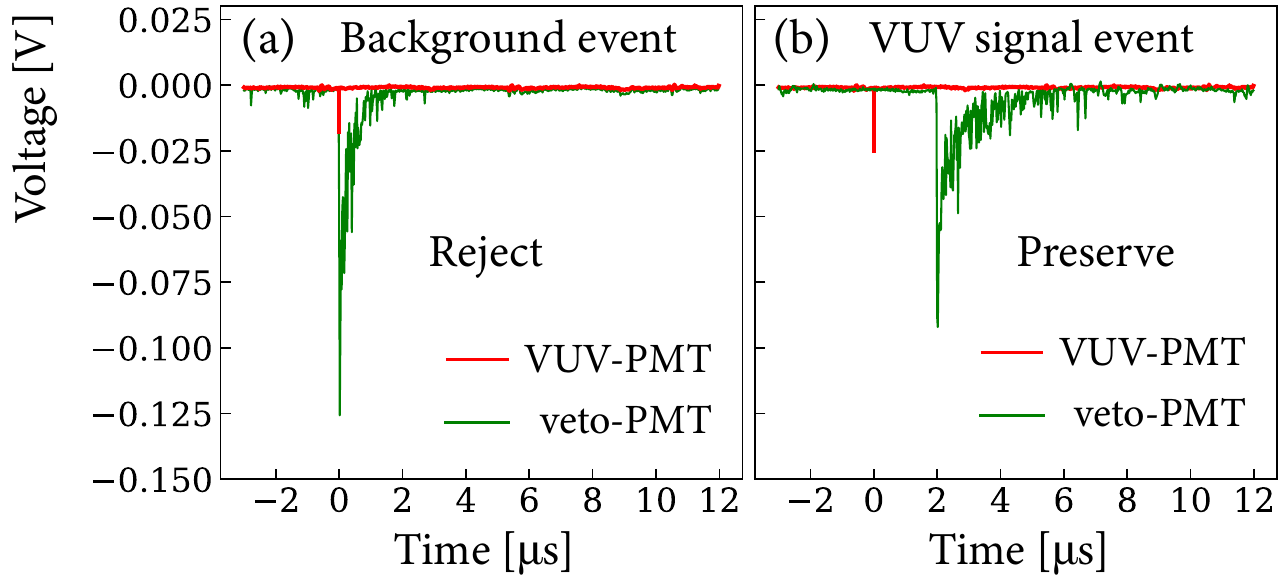}
\caption{Example waveforms of a background event and a VUV signal event. (a) Background event: a radioluminescence light burst detected by the veto-PMT coincident with the photon detection in the VUV-PMT. (b) VUV signal event: no radioluminescence light burst is detected by the veto-PMT at the instant when the VUV-PMT registers a photon.
}
\label{fig:background_signal_waveform}
\end{figure}
To reject Cherenkov-radiation-induced false signals in the VUV-PMT, an additional UV-sensitive photomultiplier tube (veto-PMT, Hamamatsu R11265-203) was installed behind the first DCM, see ``Fig.~1\,(c)''. The veto-PMT monitors STE light burst in radioluminescence, as shown by the black curves in Fig.~\ref{fig:spectra_efficiency}\,(a). The quantum efficiency of the veto-PMT is shown in Fig.~\ref{fig:spectra_efficiency}\,(b). Together with the VUV-PMT, the two detector form an anti-coincidence scheme that effectively suppresses radioluminescence-induced background events.

During the measurements, the electronic signals generated by the PMT detectors were first amplified and then recorded as waveforms using the digital oscilloscope. The VUV-PMT waveform typically exhibits a single-photon pulse, whereas the veto-PMT waveform reveals the radioluminescence light burst, showing a large peak, see Fig.~\ref{fig:background_signal_waveform}. The background rejection was achieved by analyzing the timing and temporal distribution of the light bursts peak in the veto-PMT waveforms.

Fig.~\ref{fig:background_signal_waveform} shows the typical waveforms recorded by the oscilloscope, with a capture duration of 15\,$\mu$s and a trigger threshold of $-0.005$\,V. In each waveform, the time origin ($t=0$) corresponds to the trigger timing from the VUV-PMT channel.

An example of a background event is shown in Fig.~\ref{fig:background_signal_waveform}\,(a), where the veto-PMT detects a prominent radioluminescence light burst coincident with the VUV-PMT trigger at $t=0$. Such events are rejected during the analysis. In contrast, a representative event corresponding to a potential isomer signal is shown in Fig.~\ref{fig:background_signal_waveform}\,(b), where no light burst is observed in the veto-PMT at the trigger time. Events of this type are retained as signal candidates.

\subsection{A.3 The background subtraction}

\begin{figure}[t]
    \centering
    \includegraphics[width=\linewidth]{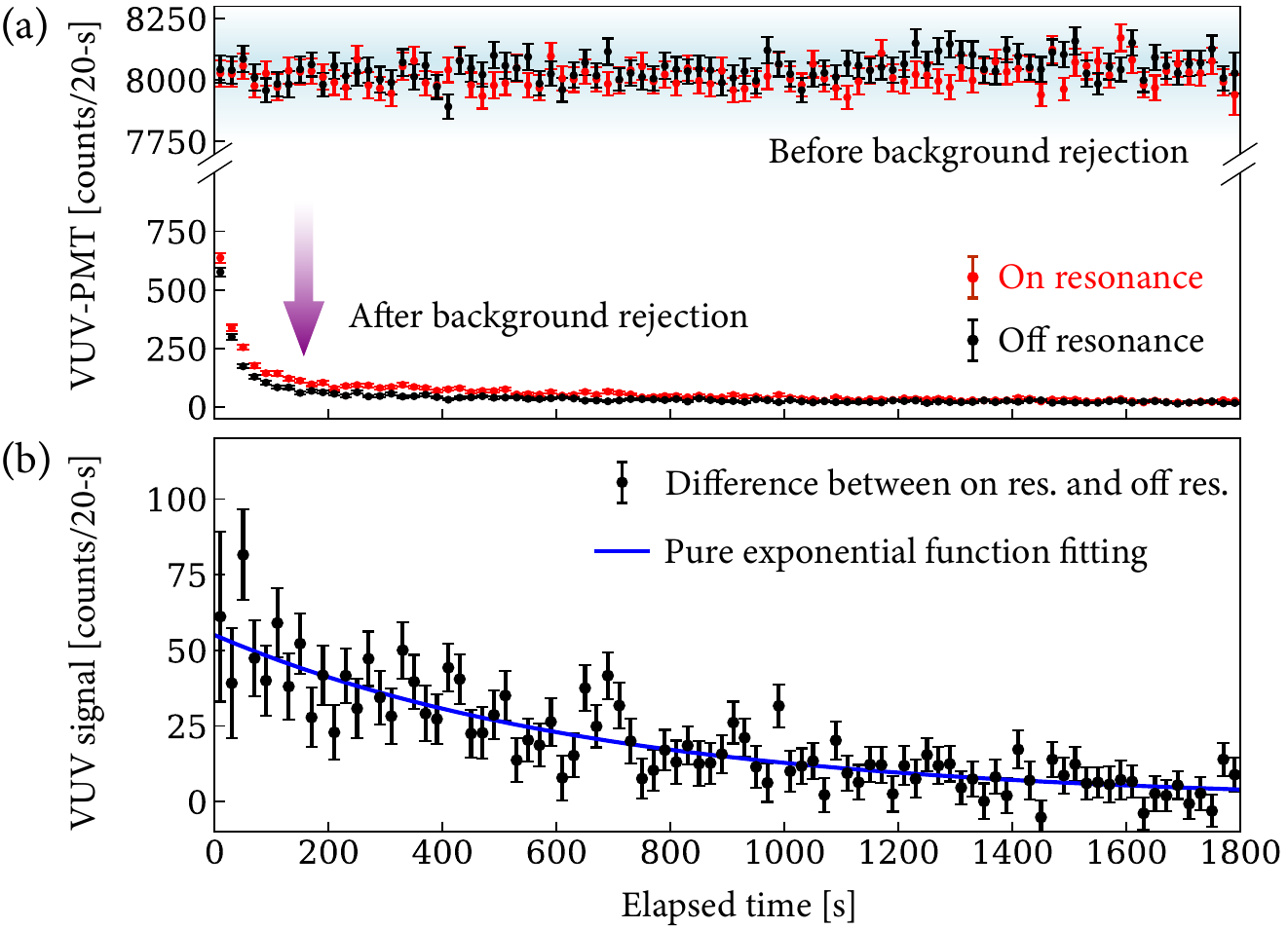}
    \caption{The background rejection and subtraction scheme. (a) Top: detector counts per 20\,s interval following target crystal irradiation with on-resonance X-rays (red) and off-resonance X-rays (black). Bottom: the corresponding results after applying the background rejection procedure. (b) The difference between the background-rejected on-resonance and off-resonance counts shown in the bottom panel of (a). }
    \label{fig:rejection_difference}
\end{figure}

In the nominal beamtime measurements, the target crystal was first irradiated with on-resonance X-rays for 300\,s, followed by a 1800\,s VUV photon detection period using the VUV-PMT. After the on-resonance measurement, the X-ray energy was detuned by $+$100\,meV to an off-resonance condition, and the same 300\,s irradiation and 1800\,s detection procedures were repeated. We define one \textit{set} as a complete cycle consisting of both an on-resonance and an off-resonance measurement.

The top part of Fig.~\ref{fig:rejection_difference}\,(a) shows the VUV-PMT count rates averaged over six sets of measurements. In this plot, the VUV-PMT counts are histogrammed according to the relative timestamp, using 20-s time bins. As shown, no discernible difference indicating the signal is observed between the on-resonance (red) and off-resonance (black) conditions. In both cases, the VUV-PMT continuously detects approximately 8000 photons per 20-s interval.

After applying the background rejection method demonstrated in A.2 (see Fig.~\ref{fig:background_signal_waveform}), the VUV-PMT count rate is significantly reduced (see the violet downward arrow), as shown in the lower part of Fig.~\ref{fig:rejection_difference}\,(a), where a clear enhancement in the on-resonance X-ray condition becomes observable. By subtracting the background-rejected detector counts obtained under the off-resonance condition from those under the on-resonance condition, we extracted the net isomer VUV signal, as shown in Fig.~\ref{fig:rejection_difference}\,(b).

Fig.~\ref{fig:rejection_difference}\,(b) shows a clear long-lifetime fluorescence signal following the cessation of X-ray pumping. The data were fitted with a single exponential decay function:
\begin{equation}
    N_{\mathrm{VUV}}(t) = N_0 \exp(-t/\tau), \label{measure_fit}
\end{equation}
as indicated by the blue line in Fig.~\ref{fig:rejection_difference}\,(b). The fit to this representative dataset yields $N_0 = 55.3(3.7)$ and a lifetime of $\tau = 685(49)$\,s. This value agrees with previous measurements of the $^{229\mathrm{m}}$Th lifetime in CaF$_2$ reported by PTB\,\&\,TU Wien~\cite{tiedau2024laser} and JILA~\cite{zhang2024frequency}. This consistency supports that, after effective background suppression and subtraction, the detected fluorescence originates from isomeric decay rather than spurious VUV contributions.

Further details about the experimental setup are provided in Ref.~\cite{guan2025method}, the data analysis procedure is described in Ref.~\cite{hiraki2024experimental}, and the isomer lifetime and the signal wavelength measurements are reported in Ref.~\cite{hiraki2024controlling}. In the following sections of this Supplemental Material, we present the measurement schemes and data analysis procedures relevant to the main manuscript.

\section{B. The measurement protocols for the observables}

\subsection{B.1 The measurement of the isomer yield $N_0$}
Currently, the most widely accepted value for the isomer natural decay lifetime in CaF$_2$ is $\tau_0 = 641(4)$\,s, as measured by the JILA research group. To enable consistent comparison of isomer yields under different experimental conditions, we fixed the lifetime parameter in Eq.~\eqref{measure_fit} to $\tau = \tau_0$. For example, the fit result for the dataset shown in Fig.~\ref{fig:rejection_difference}\,(b) yields $N_0 = 58.0(2.3)$. More specifically,
\begin{equation}
    N_0(t_{\mathrm{pump}} = 300\,\mathrm{s},\, T = 36\,^{\circ}\mathrm{C},\,\tau=\tau_0) = 58.0(2.3), \label{NtpumpT}
\end{equation}
where $t_{\mathrm{pump}} = 300\,\mathrm{s}$ denotes the X-ray irradiation time used for pumping the isomeric states, and $T = 36\,^{\circ}\mathrm{C}$ indicates that the measurement was conducted at room temperature.  

As introduced in End Matter and this Supplement Material, the obtained $N_0$ in Eq.~\eqref{NtpumpT} is proportional to the total pumped isomer states $N_{\mathrm{iso}}$. For simplicity, we defined the detection factor $\epsilon$, such that $N_0=\epsilon N_{\mathrm{iso}}$. Therefore, in this work, we use the direct observable $N_0$ as a proxy of isomer yield $N_\mathrm{iso}$. 

To investigate the influence of temperature on the isomer yield, we conducted a series of measurements during one beamtime session, setting $t_{\mathrm{pump}} = 900\,\mathrm{s}$ and systematically varying the crystal temperature from $-50\,^{\circ}\mathrm{C}$ to $-150\,^{\circ}\mathrm{C}$ in 10-degree steps. The extracted $N_0$ values, obtained following the data analysis procedures introduced in Section A.3 and Eq.~\eqref{NtpumpT}, are summarized in Tab.~\ref{tab:N0T}. 

\begin{table}[t]
\centering
\caption{The dependence of the isomer yield ($N_0$) on the crystal temperature by irradiating the crystal for 900\,s at various temperature conditions. The target crystal was cooled from -50\,$^\circ$C to -150\,$^\circ$C in 10 degrees step. This result is illustrated as the cyan error-bar points in ``Fig.~3\,(d)''.}
\begin{tabular}{cccccc}
\hline
\hline
~~\#~~ & ~~$t_{\mathrm{pump}}$~~ & ~~$T$~~& ~~Sets~~ & ~~$N_0$ mean~~ & $N_0$~uncertainty \\
\hline 
1 & 900\,s &36\,$^\circ$C &2& 40& 4   \\
2 & 900\,s &-50\,$^\circ$C&2& 62.4& 5.6  \\
3 & 900\,s &-60\,$^\circ$C&2& 65.6& 5.8  \\
4 & 900\,s &-70\,$^\circ$C&2& 55.2& 6.2  \\
5 & 900\,s &-80\,$^\circ$C&2& 44.2& 6.2  \\
6 & 900\,s &-90\,$^\circ$C&2& 46.1& 6.6  \\
7 & 900\,s &-100\,$^\circ$C&2& 52.3& 7.2  \\
8 & 900\,s &-110\,$^\circ$C&2& 59.8& 8.2  \\
9 & 900\,s &-120\,$^\circ$C&2& 65.04& 10.0  \\
10 & 900\,s &-130\,$^\circ$C&2& 79.4& 12.1  \\
11 & 900\,s &-140\,$^\circ$C&2& 102.6& 13.8  \\
12 & 900\,s &-150\,$^\circ$C&2& 113.4& 15.4  \\
\hline
\hline
\end{tabular}
\label{tab:N0T}
\end{table}
The measured temperature dependence of the isomer yield, summarized in Tab.~\ref{tab:N0T}, is plotted in ``Fig.~3\,(d)''. The data demonstrate that the isomer yield increases at lower temperatures, with an anomalous behavior observed at $T = -60\,^{\circ}\mathrm{C}$. 

Meanwhile, following the 900\,s irradiation, the afterglow fluorescence emitted from the target crystal was also captured by the veto-PMT. In the experimental setup, unbiased trigger events—independent of the VUV-PMT triggers—were used to extract the afterglow-induced waveforms. The integrated  afterglow waveform represents the afterglow light yield; four afterglow examples are shown in ``Fig.~4\,(a)''. For comparison, both the afterglow yields and the isomer yields measured at different temperature are plotted in ``Fig.~4\,(b)''.

\subsection{B.2 The measurement of the quenched lifetime $\tau_{\mathrm{ir}}$}

\begin{table}[t]
\centering
\caption{The dependence of the isomer yield ($N_0$) on the irradiation time $t_{\mathrm{pump}}$. The target crystal was sustained at room temperature (36\,$^\circ$C) while the irradiation time varied from 5\,s to 600\,s. This result is illustrated as the blue error-bar points in ``Fig.~3\,(a)''.}
\begin{tabular}{cccccc}
\hline
\hline
~~~\#~~ & ~~$t_{\mathrm{pump}}$~~ & ~~$T$~~& ~Sets~ & ~~~$N_0$ mean~~~ & $N_0$~uncertainty \\
\hline
1& 5 & 36\,$^\circ$C & 1 &4.4 &5.6 \\
2& 10 & 36\,$^\circ$C & 1 &20.4 &5.8 \\
3& 20 & 36\,$^\circ$C & 1 & 14.4 &5.9 \\
4& 30 & 36\,$^\circ$C & 1 &17.4 &6.0 \\
5& 40 & 36\,$^\circ$C & 1 &22.1 &5.9 \\
6& 50 & 36\,$^\circ$C & 1 &28.6 &6.0 \\
7& 60 & 36\,$^\circ$C & 1 &17.2 & 6.1\\
8& 80 & 36\,$^\circ$C & 1 & 40.4 &6.1 \\
9& 100 & 36\,$^\circ$C & 1 &40.5 & 6.1\\
10& 150 & 36\,$^\circ$C & 1 &42.3 &6.1 \\
11& 300 & 36\,$^\circ$C & 1 &40.1 &6.1 \\
12& 600 & 36\,$^\circ$C & 1 &41.3 & 6.2\\
\hline
\hline
\end{tabular}
\label{tab:Tpump}
\end{table}

To investigate the isomer population process during X-ray irradiation, we conducted a series of measurements as summarized in Tab.~\ref{tab:Tpump}. The measurements were performed at room temperature, with the X-ray irradiation time ($t_{\mathrm{pump}}$) on the target crystal varied from 5\,s to 600\,s. For each irradiation time, one full measurement set was carried out. The extracted isomer yields and their uncertainties are listed in the last two columns of Tab.~\ref{tab:Tpump} and plotted in ``Fig.~3\,(a)''. Additionally, the combined VUV detection signals from entries 11 and 12 in the table are shown in ``Fig.~3\,(c)''. 

Meanwhile, we performed similar $t_{\mathrm{pump}}$-dependent measurements at different temperatures. The results for the $T = -164\,^{\circ}\mathrm{C}$ condition are also plotted in ``Fig.~3\,(a) and (c)'', where the influence of temperature on the saturation behavior of the isomer yield is evident.

The $t_{\mathrm{pump}}$-dependent measurements at different temperature conditions are summarized in Tab.~\ref{tab:tauIR}, where the column labeled “Points” indicates the number of measurements conducted at each temperature. For each measurement, the isomer population buildup during X-ray irradiation was fitted using the following function:
\begin{equation}
    N(t) = N_0 \left[1 - \exp\left(-t/\tau_{\mathrm{ir}}\right)\right],
\end{equation}
where $\tau_{\mathrm{ir}}$ represents the characteristic population time constant during irradiation. The physical meaning of this fitting function was reported in Ref.~\cite{hiraki2024controlling}, and the derivation process is provided in the End Matter. Two fitting examples for $T = 36\,^{\circ}\mathrm{C}$ and $T = -164\,^{\circ}\mathrm{C}$ are illustrated in ``Fig.~3\,(a)''. 

The fitting results for all temperature conditions are listed in the last two columns of Tab.~\ref{tab:tauIR} and plotted in  ``Fig.~3\,(b)''. This figure clearly shows that the isomer population time constant during X-ray irradiation, $\tau_{\mathrm{ir}}(T)$, exhibits a strong temperature dependence: as $T$ decreases, $\tau_{\mathrm{ir}}(T)$ increases (with the exception of an anomalous point near $-60\,^{\circ}\mathrm{C}$).  The main trend observed in ``Fig.~3\,(b)'' is fitted using the following function:
\begin{equation}
    \tau_{\mathrm{ir}}(T) = \frac{\tau_0}{1 + \beta \exp(-E_a/kT)},
\end{equation}
where $\beta$ and $E_a$ are fitting parameters, and $k$ is the Boltzmann constant. The derivation of this fitting function is provided in the End Matter.

\begin{table}[t]
\centering
\caption{The dependence of the isomer lifetime during X-ray irradiation  on the target crystal temperature. The $\tau_{\mathrm{ir}}$ values are retrieved from the measurement protocols exampled in Tab.~\ref{tab:Tpump}. This result is illustrated as the green error-bar points in ``Fig.~3\,(b)''.}
\begin{tabular}{ccccc}
\hline
\hline
~~~\#~~ & ~~~$T$~~~  & ~~~ Points ~~& ~~ $ \tau_{\mathrm{ir}}$ mean~~ &~~ $\tau_{\mathrm{ir}}$ uncertainty~~  \\
\hline 
1&36.5\,$^\circ$C &10& 60.9 & 19.5\\
2&36\,$^\circ$C  &12& 45.9  & 21.8 \\
3&-15\,$^\circ$C &8& 106.6 & 34.5\\
4&-38\,$^\circ$C &9& 122.0 &36.0 \\
5&-60\,$^\circ$C &11& 215.2 &57.4 \\
6&-80\,$^\circ$C &9& 139.2 & 48.3 \\
7&-100\,$^\circ$C &10& 172.0 & 70.0 \\
8&-120\,$^\circ$C &10& 194.2 &82.3 \\
9&-164\,$^\circ$C &13& 318.1 &64.6 \\
\hline
\hline
\end{tabular}
\label{tab:tauIR}
\end{table}

\section{C. The cooling experiments}

\subsection{C.1 Cooling method}

The crystal was cooled via thermal conduction from a stainless-steel mount to the cold head of a Cryo-cooler (setup detailed in Ref.~\cite{guan2025method}). During cooling, photon-absorption by an \textit{ice-layer} was observed, attributed to residual gas molecules in the vacuum vessel condensing on the cold crystal surface and attenuating emitted photons. This behavior is consistent with previous observations in laser excitation experiments by PTB and TU Wien~\cite{tiedau2024laser,Tiedau_PRL_SM}.

\subsection{C.1 Ice-layer suppression}
The setup described in Ref.~\cite{guan2025method} enables cooling of the crystal to $\mathrm{-190\,^\circ C}$. However, at lower temperatures, ice-layer absorption becomes increasingly significant. In particular, below $\mathrm{-100\,^\circ C}$, the ice can strongly attenuate both signal photons and Cherenkov background light.

To suppress ice-layer absorption, we developed a cold copper shield surrounding the crystal to prevent gas molecules from condensing on its surface. Although the copper shield is not shown in the Th-setup in ``Fig.~1'', all data presented in this work was acquired with the shield installed.

\subsection{C.2 veto-PMT waveform deformation}
When varying the crystal temperature to study isomer quenching, we also encountered side effects on the luminescence properties. As discussed in the main manuscript, there is a strong correlation between isomer quenching and luminescence behavior at different temperatures. A detailed investigation of the luminescence, which supports the proposed quenching model, is ongoing and beyond the scope of this Supplemental Material. Here, we emphasize that temperature variation alters the properties of the light burst in radioluminescence. These light burst are detected by the veto-PMT waveform, and the deformation of these waveforms adds complexity to the background rejection algorithm.

As shown in Fig.~\ref{fig:background_signal_waveform}, waveform (a) was recorded at room temperature, while waveform (b) was taken at $\mathrm{-60\,^\circ C}$. At lower temperatures, the peak amplitude decreases and the tail becomes elongated. This behavior reflects the suppression of \textit{thermal quenching} of the STE in CaF$_2$~\cite{williams1990self, mikhailik2006scintillation}, leading to increased light yield and longer STE lifetime. The deformation becomes more pronounced with decreasing temperature, and below $\mathrm{-165\,^\circ C}$, the light burst peaks are smeared out, rendering background rejection analysis infeasible. Consequently, data acquisition was halted at this temperature.

\end{document}